\documentclass[pra,aps,floatfix,showpacs,tightenlines,superscriptaddress,amsmath,amssymb,showkeys,10pt,nofootinbib]{revtex4-2}

\usepackage{slashed}
\usepackage{xcolor}
\usepackage{graphicx}

\usepackage{hyperref}

\newcommand{\be}{\begin{equation}}\newcommand{\ee}{\end{equation}}
\newcommand{\bea}{\begin{eqnarray}}\newcommand{\eea}{\end{eqnarray}}
\newcommand{\brr}{\begin{array}}\newcommand{\err}{\end{array}}
\newcommand{\bit}{\begin{itemize}}\newcommand{\eit}{\end{itemize}}
\newcommand{\ben}{\begin{enumerate}}\newcommand{\een}{\end{enumerate}}

\newcommand{\bbm}{\begin{bmatrix}}\newcommand{\ebm}{\end{bmatrix}}
\newcommand{\ba}{\begin{array}}
\newcommand{\ea}{\end{array}}

\newtheorem{mydef}{Definition}
\newtheorem{Lemma}{Lemma}
\newcommand{\bd}{\begin{mydef}} \newcommand{\ed}{\end{mydef}}
\newcommand{\bthe}{\begin{theorem}} \newcommand{\ethe}{\end{theorem}}
\newcommand{\ble}{\begin{Lemma}} \newcommand{\ele}{\end{Lemma}}

\newcommand{\dr}{\mathrm{d}}

\def\ha{\frac{1}{2}}

\def\ph{\varphi}

\def\lf{\left}

\def\ri{\right}
\def\al{\alpha}\def\ga{\gamma}

\def\1{{_{1}}}\def\2{{_{2}}}

\def\noHe0{:\;\!\!\;\!\!:H_e(0):\;\!\!\;\!\!:}
\def\noHm0{:\;\!\!\;\!\!:H_\mu(0):\;\!\!\;\!\!:}

\def\lf{\left}

\def\ri{\right}

\def\al{\alpha}\def\ga{\gamma}

\def\1{{_{1}}}\def\2{{_{2}}}

\begin{document}

\title{Ordering Kinetics of the two-dimensional voter model with long-range interactions}

\author{Federico Corberi}
\email{fcorberi@unisa.it}
\affiliation{Dipartimento di Fisica, Universit\`a di Salerno, Via Giovanni Paolo II 132, 84084 Fisciano (SA), Italy}
\affiliation{INFN Sezione di Napoli, Gruppo collegato di Salerno, Italy}

\author{Luca Smaldone}
\email{lsmaldone@unisa.it}
\affiliation{Dipartimento di Fisica, Universit\`a di Salerno, Via Giovanni Paolo II 132, 84084 Fisciano (SA), Italy}
\affiliation{INFN Sezione di Napoli, Gruppo collegato di Salerno, Italy}

\begin{abstract}
We study analytically the ordering kinetics of the two-dimensional long-range voter model on a two-dimensional lattice, where agents on each vertex take the opinion of others at distance $r$ with probability $P(r) \propto r^{-\al}$. The model is characterized by different regimes, as $\al$ is varied. For $\al > 4$ the behaviour is similar to that of the nearest-neighbor model, with the
formation of ordered domains of a typical size growing as $L(t) \propto \sqrt{t}$, until consensus is reached in a time of the order of $N\ln N$, with $N$ being the number of agents. Dynamical scaling is violated due to an excess of interfacial sites whose density decays as slowly as $\rho(t) \propto 1/\ln t$.
Sizable finite-time corrections are also present, which are absent in the case of nearest-neighbors interactions. For $0<\al \leq 4$ standard scaling is reinstated, and the correlation length increases algebraically as $L(t)\propto t^{1/z}$, with $1/z=2/\al$ for $3<\al<4$ and $1/z=2/3$ for $0<\al<3$. In addition, for $\al \le 3$, $L(t)$ depends on $N$ at any time $t>0$. Such coarsening, however, only leads the system to a partially ordered metastable state where correlations decay algebraically with distance, and whose lifetime diverges in the $N\to \infty$ limit. In finite systems consensus is reached in a time of order $N$ for any $\al <4$.

\end{abstract}

\maketitle
\section{Introduction}
The voter model was firstly introduced in the study of genetic correlations \cite{Kimura1964,1970mathematics}. Later on its basic properties were derived in Refs. \cite{Clifford1973,Holley1975} and widely studied through the years \cite{Clifford1973,Holley1975,liggett2004interacting,Theodore1986,Scheucher1988,PhysRevA.45.1067,Frachebourg1996,Ben1996,PhysRevLett.94.178701,PhysRevE.77.041121,RevModPhys.81.591}, with application to various disciplines \cite{Zillio2005,Antal2006,Ghaffari2012,CARIDI2013216,Gastner_2018,RevModPhys.81.591,Redner19}.
The informing idea is quite simple: at each point of a lattice an agent can express one of two possible choices, say ``left'' and ``right'' or, equivalently $+1$ or $-1$. Because of the similarity to the Ising model, the term ``spins'' is also used for such agents. In the original nearest-neighbor (NN) version, an agent takes the status of a randomly chosen NN, so that the probability of the spin-flip is proportional to the fraction of opposite NN agents surrounding it. 
In arbitrary space dimension, this dynamical rule does not respect detailed balance, and hence the voter model, although perhaps adequate for the description of some social system and other phenomena, is not fitted to describe, even at an elementary level, the thermodynamic properties of physical substances such as magnetic materials.
However, regardless of its application domain, the interest in the voter model is mainly due to the fact it can be exactly solved for every number of space dimensions \cite{PhysRevA.45.1067,Frachebourg1996}, which is not true for the Ising model. 
Then, the critical behavior of a broad variety of models belonging to the universality class of the voter one can be easily determined~\cite{PhysRevLett.87.045701}. 
Furthermore, despite the differences between the two models, it has been sometimes argued that the analytical solution of the voter model could shed some light on the properties of the less tractable
Ising one. In a sense, in passing from the latter to the former we {\it trade detailed balance against exact solubility}~\cite{Scheucher1988}. 

Through the years, many variants of the original model have been proposed, in order to adapt it to explain different situations \cite{Mobilia2003,Vazquez_2004,MobiliaG2005,Dall'Asta_2007,Mobilia_2007,Stark2008,Castellano_2009,Moretti_2013,Caccioli_2013,HSU2014371,PhysRevE.97.012310,Gastner_2018,Baron2022}. In particular, in a recent paper~\cite{corberi2023kinetics} the ordering kinetics of the one-dimensional voter model in the presence of long-range interactions have been analytically studied. There, the interaction probability between two agents at distance $r$ was taken to be of the form $P(r) \propto r^{-\al}$. Various regimes of $\al$ were analyzed, displaying a rich and diverse structure. For $\al>3$ the model behaves as the NN ($d=1$) one, with a typical size of the domains growing as $\sqrt{t}$ until consensus is reached. Moreover, the correlation function presents a dynamical scaling behavior, with corrections at large distances. For smaller $\al$ a breaking of scaling was found, and lowering $\al$ to $\al \le 2$ the formation of partially ordered stationary states was discovered, similarly to what is known for the mean-field case, corresponding to $\al=0$. 

Besides the interest in the voter model itself, the case with extended interactions can also be framed into the more general topic of long-range
statistical models far from equilibrium, to which an increasing interest has been devoted in recent years, concerning phase-ordering kinetics~     \cite{PhysRevE.49.R27,PhysRevE.50.1900,PhysRevE.99.011301,Corberi_2017,Corberi2019JSM,PhysRevE.103.012108,Corberi2021SCI,CORBERI2023113681,Corberi2023PRE,PhysRevE.102.020102} and other aspects~\cite{campa2009statistical,book_long_range,DauRufAriWilk}.

In this work we extend the analysis of the ordering dynamics of the long-range voter model presented in Ref. \cite{corberi2023kinetics}, by studying the evolution of the two-dimensional case, where quite different results are observed. A brief summary of what we find follows. For $\al >4$ the model's behaviour is akin to that of the NN model: a correlation length grows as
$L(t)\propto \sqrt t$ and there are violations of dynamical scaling due to the presence of a large amount of interfacial agents, whose number decays as $\rho(t)=1/\ln t$. Due to that, the consensus time grows as $T \propto N \ln N$ with the number $N$ of agents, as in the NN case. In addition, important finite-time corrections can be explicitly computed for $x \equiv r/L(t)$ larger than $x^* \propto \sqrt{(\al-4) \ln t}$, at variance with the NN case.  For $\al  \leq 4$  the system presents non-trivial stationary states, with the correlations decaying with distance $r$ as $r^{-(4-\al)}$ or as $r^{-\al}$ for $2<\al \le 4$ and $0\le \al \le 2$, respectively. Such stationary state is approached through a coarsening regime with $L(t) \propto t^{\frac{2}{\al}}$ for $3 \leq \al \leq 4$ and $L(t) \propto t^{\frac{2}{3}}$ for $0 < \al <3$. For $\al \le 3$, it is also found that $L(t)$ monotonically increases with $N$ at any $t>0$ and diverges in the the thermodynamic limit.

The paper is organized as follows: In Sec.~\ref{secmodel} we define the voter model
and derive the evolution equation for the
equal-time correlation function, the fundamental observable from which most of the dynamical properties can be studied. 
Then, in the following Secs.~\ref{secalgt4}, \ref{secalin24} and \ref{secallt2}, we separately study the kinetics of the model in the $\alpha$-sectors $\alpha>4$, $2<\alpha \le 4$, and $0\le \al \le 2$, where different 
dynamical properties are found.
In Sec.~\ref{secsizedep} we determine the
dependence of the coarsening growing length 
on the system size.
A final discussion and some future research perspectives are provided in the last Sec.~\ref{secconcl}.


\section{The model} \label{secmodel}

In the voter model a set of binary variables (spins), assuming the value $S_i=\pm 1$, is located on a $d$-dimensional lattice and interacts with a probability $P(\ell)=\frac{1}{Z}\,\ell^{-\alpha}$, where $\ell$ is the distance between two of them. $Z$ is a normalization that, for $\alpha \le d$, depends on the number $N$ of spins. In general, distances are non-integer
numbers: for instance in the case of a square lattice in $d=2$, $r=1, \sqrt{2}, 2, \ldots$.
Hence it is useful to introduce an integer index $p$, indicating {\it proximity}, such that $p=1$ means NN, $p=2$ next-NN and so on.
Then 
\be
Z=\sum_pn_p\,\ell_p^{-\al},
\label{eqZ} 
\ee
where $n_p$ is the number of lattice sites at {\it proximity} $p$, and $\ell_p$ is the distance between two $p$-neighboring sites. In Eq.~(\ref{eqZ}) $p$ runs from 1 to the maximum proximity number in the considered lattice.
To ease the notation this will always be implicitly understood in any $p$-summation, unless differently stated. Regarding $n_p$, it is clear that it is a multiple of $2d$ and that it is an irregularly increasing function of $p$, fluctuating around
its continuum approximation $n_p=\Omega _{d-1} l_p^{d-1}$, with $\Omega _d$ being the surface of a unit $d$-dimensional sphere. 

The probability to flip a spin $S_i$ is
\be
w(S_i)=\frac{1}{2N} \, \sum _p P(\ell_p) \sum _{|k-i|=\ell_p}(1-S_iS_k) \, ,
\ee
where $k$ are the $n_p$ sites $p$-neighbors of $i$.
Our aim is to find the correlation functions  $C(r,t) =\langle S_i(t)S_{j}(t)\rangle$, where $r=|i-j|$ is the distance between the $i$-th and the $j$-th sites. Following Ref.~\cite{Glauber} one has
$\frac{d}{dt}\langle S_{i_1}S_{i_2}\cdots S_{i_n}\rangle=-2\langle S_{i_1}S_{i_2}\cdots S_{i_n}\cdot \sum _{m=1}^n w(S_{i_m})\rangle$
which, for $n=2$, provides
\begin{eqnarray}
\frac{d}{dt}\langle S_i(t)S_j(t)\rangle&=&-2\langle S_i(t)S_j(t)\rangle+\sum _pP(\ell_p)\left [\sum_{|k-i|=\ell_p}\langle S_j(t)S_k(t)\rangle
+\sum _{|q-j|=\ell_p}\langle S_i(t)S_q(t)\rangle \right ] \, ,
\label{eqc0}
\end{eqnarray}
 where time is measured in units of $N$ elementary moves, i.e. in Monte Carlo steps.
 
 \begin{figure}[htbp]
 	\centering
 	\includegraphics[width=0.5\textwidth]{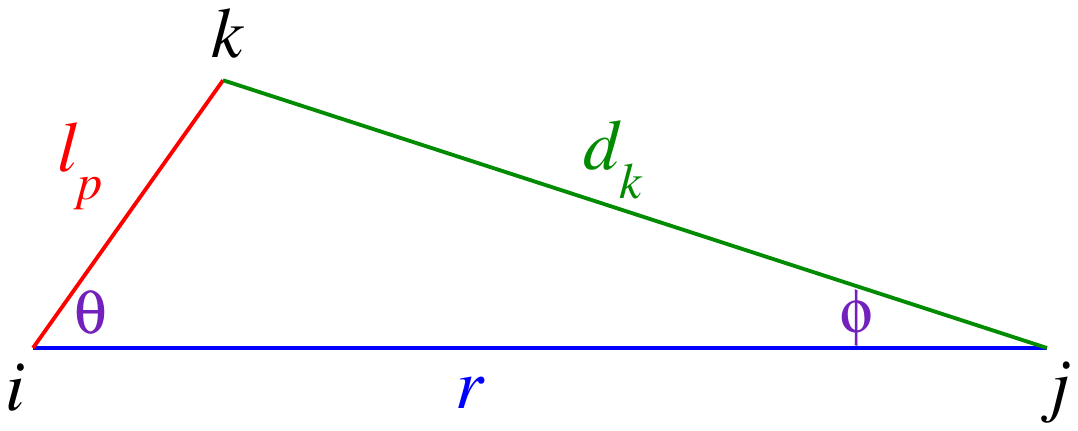}
 	\caption{The points $i,j,k$ and their distances.}
 	\label{fig_distances}
 \end{figure}
 
Then, Eq.~\eqref{eqc1}, can be written as
 \begin{eqnarray}
\dot C (r,t) &=&-2 C (r,t) +2\sum _pP(\ell_p) \sum _{k=1}^{n_p}C ([\![d_k(r,\ell_p)]\!],t)\, ,
\label{eqc1}
\end{eqnarray}
where the dot is a time derivative and the double square-brackets indicate that we are using periodic boundary conditions. In other words, the topology of the system becomes that of a torus, which is obtained by gluing together the edges of the lattice. This means that
\be
[\![n]\!]=\left \{ \begin{array}{ll}
n, & \mbox{if }n\in {\cal D} \\
{\cal M}(n), & \mbox{if } n\notin {\cal D},
\end{array} \right .
\ee
where ${\cal D}$ is the set of all possible distances on a quarter of the lattice, and ${\cal M}(n)$ is the shorter distance computed moving through
the boundary. The various distances entering Eq.~(\ref{eqc1}) are sketched in Fig.~\ref{fig_distances}.
Let us remark that Eq.~(\ref{eqc1}) is exact
in any $d$.
Its solution in $d=2$ will be obtained, by approximating in different ways the sums contained on the r.h.s., in the following sections. 

A dynamical correlation length can be extracted from $C$ as 
\be
L(t)=\frac{\sum _{p} n_p \, r_p\,C(r_p,t)}{\sum _{p} n_p C(r_p,t)} \, .
\label{eqL}
\ee
Notice that, generally speaking, $L(t)$ depends on $N$, however we do not explicitly indicate such dependence to ease the notation. 

In the following we will be interested in the 
ordering kinetics of the model when it is
prepared in a fully disordered initial condition, i.e. $P(S_i)=\frac{1}{2}\delta _{S_i,1}+\frac{1}{2}\delta _{S_i,-1}, \, \forall i$,
with $C(r,t=0)=\delta_{r,0}$.
Typical behaviors of $L(t)$ for such process on the 2-$d$ square lattice, which will be further commented on in the following sections, can be seen in Fig.~(\ref{fig_length}). This and the following figures of this
article are obtained by numerically solving Eq.~(\ref{eqc1})
and represent, therefore, exact results (apart, clearly, from numerical and/or discretisation errors) \cite{supmat}. Once $C(r,t)$ has been obtained in this way,
$L(t)$ is computed by means of its definition~(\ref{eqL}). Generally, one sees in Fig.~\ref{fig_length} that there is an initial coarsening stage 
in which $L(t)$ increases (typically algebraically in time), until saturation to a final value of the order of $10^2$ is attained. This is a finite-size effect setting-in when $L(t)$ becomes comparable to the system size and, hence, can be pushed to larger and larger values of $L$ in the thermodynamic limit $N\to \infty$.

\begin{figure}[htbp]
	\centering
		\includegraphics[width=0.8\textwidth]{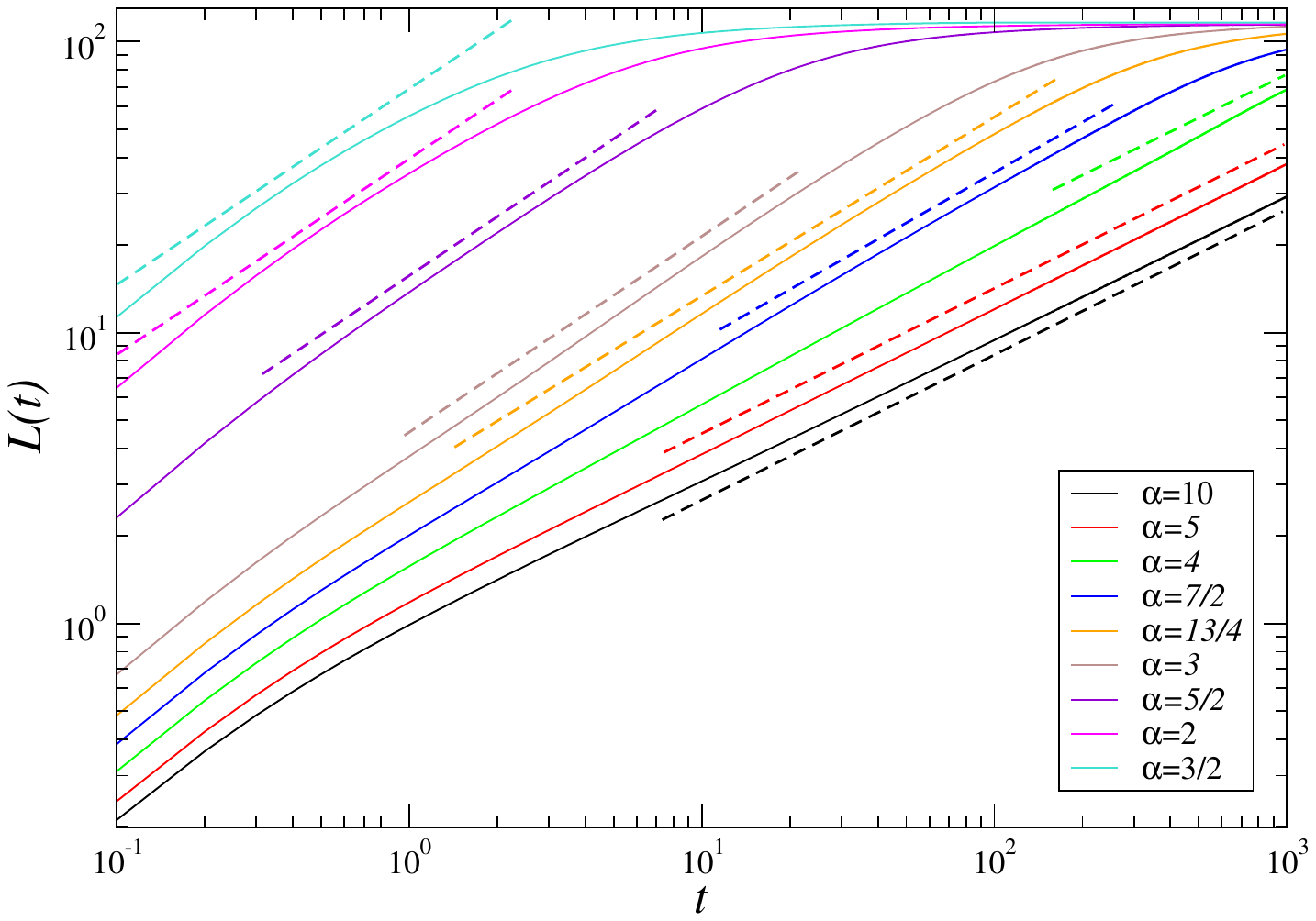}
	\caption{$L(t)$ is plotted against time on double-logarithmic axes, for different values of $\al$, see legend. System size is $N=301^2$. The straight dashed lines are the analytic predictions~(\ref{growthagt4},\ref{growthain23},\ref{growthalt3}).}
	\label{fig_length}
\end{figure}

In the rest of this article we will study the model introduced insofar, specializing to the two-dimensional
case. In order to do this, we have to separately consider different regimes for $\al$.

\section{Case \boldmath{$\alpha >4$}} \label{secalgt4}

Since Eq.~(\ref{eqc1}) cannot be routinely solved in a simple way,
we proceed by guessing the form of the solution and then verifying
the consistency of the assumption at the end of the calculation,
benchmarking also with the results of a numerical integration of 
the same equation.

For sufficiently large values of $\alpha$ -- which will turn out to be
$\alpha >4$ -- we expect a form of $C(r,t)$ analogous to the
one known for the NN model~\cite{PhysRevA.45.1067}:
\be
C(r,t) \ = \ \frac{1}{\ln t} \, f\left (\frac{r}{L(t)}\right ),
\label{scalAnsatzSimple}
\ee
where $L(t)$ is a function of time to be determined, which reduces to 
$L(t)\propto t^{1/2}$ in the NN case (i.e. for $\alpha \to \infty$).

Substituting the condition~\eqref{scalAnsatzSimple} into Eq.~\eqref{eqc1}, we get
\be \label{diffeq}
    \frac{f(x)}{t\ln t}+
    \frac{\dot{L}(t)}{L(t)}\, x \, f^{'}(x) \ = \ 2 \lf[ f(x) -\sum _p P(\ell_p) \sum _{k=1}^{n_p}f\lf(\frac{[\![d_k(r,\ell_p)]\!]}{L(t)}\ri)\ri] \, ,
\ee
where $f'$ means a derivative with respect to $x \ \equiv \ r/L(t)$. 
Notice that the form~(\ref{scalAnsatzSimple})) implies that the argument $x$ of the function $f$ tends to become a continuous variable when $L(t)$ increases more and more. So we are allowed to use a continuum approach with respect to this variable. Enforcing this fact we can Taylor expand the argument of the sum on the r.h.s. of the above equation around $\ell=0$. We argue now that the expansion can be truncated to the lowest (significative) order. Heuristically, this is expected if 
$P(\ell)$ decays sufficiently fast (which will be shown to occur for $\alpha >4$) to make only the terms with small $\ell$ relevant. A more precise statement of this concept follows:
For small enough $x$, i.e. smaller than a certain value $x^*$ that will be determined later (see Eq.\eqref{xstar}, below), the relevant contribution to the sum on the r.h.s. of Eq.~\eqref{diffeq} comes from the points $d_k \approx r$, i.e. for small $l_p$.
Indeed, as one can verify {\it a posteriori} (see Eq.~(\ref{formf})), for small $r$, $f$ vary much less than $P$ which,
in turn, decays sufficiently fast as to make 
relevant only the contributions  
produced at small $\ell$. 
We now proceed with the Taylor expansion of $f$ around $\ell=0$ in the sum. In doing that, the first order term cancels, as it can be seen 
employing a continuum approximation, namely replacing the sum on the r.h.s. of Eq.~(\ref{diffeq})
with an integral as
\be
\sum _p P(\ell_p) \sum _{k=1}^{n_p}f \lf(\frac{[\![d_k(r,\ell_p)]\!]}{L(t)}\ri) \ \to  \ \int \!\! \dr \ell \,\ell \, P(\ell) \int^{2 \pi}_0 \!\! \dr \theta \, f \lf(\frac{d(r,\ell, \theta)\!}{L(t)}\ri) \, , 
\label{discrcont}
\ee
where $\theta$ is the angle between $\vec r$, namely the segment $i-j$ (see Fig.~\ref{fig_distances}), and
$\vec \ell _p$, namely the segment $i-k$, and
\be
d(r,\ell, \theta) \ = \ \sqrt{r^2+\ell^2-2 \ell r \cos \theta} \ \approx \ r-\ell \cos \theta \, ,
\label{eqd}
\ee
where the last relation holds for $\ell \ll r$. 
Notice that we have written $d(r,\ell,\theta)$ in Eq.~(\ref{discrcont}),
instead of $[\![d(r,\ell,\theta)]\!]$, because we are working for small $d$.
Therefore, the first order term in the Taylor expansion reads
\be
-f'(x)\int \!\! \dr \ell \, \ell^2 \, P(\ell)  \int^{2 \pi}_0 \!\! \dr \theta \, \frac{\cos \theta}{L(t)} \ = \ 0 \, .
\ee
Then, upon Maclaurin expanding $f$ on the r.h.s. of Eq.~(\ref{diffeq}) up to second order in the small quantity $\ell$, we arrive at
\be \label{diffeqm}
 \frac{f(x)}{t\ln t}+
\frac{\dot{L}(t)}{L(t)} \,x \, f^{'}(x) \ 
   = \  - \sum _p P(\ell_p) \sum _{k=1}^{n_p} \lf[ f^{''}(x) (d_k-r)^2 \, L^{-2}(t) + x f^{'}(x) \ph^2_k \ri] \,  \, ,
\ee
where $\varphi_k$ is the angle between $\vec r$ and $\vec d_k$, namely the segment $j-k$ (see Fig.~\ref{fig_distances}).

Using the law of sines, it is $\sin \varphi _k=\frac{\ell}{d_k}\,\sin \theta _k \simeq \frac{\ell}{r}\,\sin \theta_k$, the last passage holding for small $\ell$. In the same limit it is also $\sin \varphi _k \simeq \varphi _k$, leading to $\varphi _k \approx \frac{\ell}{r} \,\sin \theta _k$. Using this fact, Eq.~\eqref{diffeqm} takes the form
\be \label{diffeq1}
 \frac{f(x)}{t\ln t}+
\frac{\dot{L}(t)}{L(t)}\, x \, f^{'}(x) \ 
  = \    \, -\lf[ {\cal J} f^{''}(x)+x^{-1}{\cal K} f^{'}(x) \ri] \,  L^{-2}(t) \, ,
\ee
 where  ${\cal J} \equiv \sum _p P(\ell_p) \sum_k \, \, (d_k(r,\ell_p)-r)^2$  and ${\cal K} \equiv \sum _p l_p^2 P(\ell_p) \sum_k \sin^2 \theta_k $. Since $d-r\approx \ell \,cos \,\theta$, after Eq.~(\ref{eqd}), in the continuum limit 
\be \label{coeff}
{\cal J} \approx {\cal K} \ \approx \ \pi\int \!\! \dr \ell\,  \ell^3 P(\ell)  \, .
\ee

Notice that such expression is only convergent for $\al >4$, which
already suggests that the solution we will arrive at in this way only holds in that range of
$\alpha$-values.

We now make the ansatz, to be verified {\it a posteriori}, that 
the term $f/(t\ln t)$ can be dropped in Eq.~(\ref{diffeq1}), because it is subdominant for large $t$, yielding
\be \label{diffeq1wl}
 \frac{ \dot{L}(t)}{L(t)} \, x \, f^{'}(x)  \ = \  -{\cal J} \lf[ f^{''}(x)+x^{-1} f^{'}(x) \ri] L^{-2}(t)\, .
\ee
The r.h.s does not explicitly depend on time. In order to have the same
property on the l.h.s. it must be
\be
L(t) \ = \ D \ t^\ha \, .
\label{growthagt4}
\ee
Then we have that for $\al >4$  the correlation growth-law is the same as in the NN 
model~\cite{Ben1996}. This behavior is 
clearly observed for any value of $\alpha >4$ in Fig.~\ref{fig_length},
where data for $L(t)$ have been obtained numerically, as already explained  
at the end of Sec.~\ref{secmodel}.
Plugging the form~(\ref{growthagt4}) in Eq.~(\ref{diffeq1wl}), an equation for $f(x)$ emerges
\be
\frac{x_0^2}{2} \, f^{''}(x)+ \lf(\frac{x_0^2}{2x}+x\ri) \, f^{'}(x) \ = \ 0 \, ,
\ee
with $x_0^2 \equiv 4 {\cal J}/D^2$.  The solutions are
\be
f(x) \ = \  c_1 \text{E}_1\left[\left (\frac{x}{x_0}\right )^2\right]+c_2\, ,
\label{solutions}
\ee
where $\text{E}_1$ is the exponential integral special function~\cite{abramowitz1965handbook}. In order to fix the constant, one should impose that the correlation function \eqref{scalAnsatzSimple} goes to 1 when $r = 1$ (the minimum distance in the lattice) and $t \gg 1$, i.e. when $x \to 0$. In such case we can use that
\be \label{e1small}
\text{E}_1\left(z \right) \ \sim \ -\ga -\ln z \, , \quad \mbox{as }z \to 0 \, , 
\ee
where $\ga$ is the Euler--Mascheroni constant. Using this asymptotic result in Eq.~(\ref{solutions}) and keeping only the dominant term for large times, one has $c_1=1$. Moreover, imposing that $C(r) \to 0$ when $r \to \infty$, $c_2=0$ is also determined. Therefore
\be
f(x) \ = \  \, \text{E}_1\left[\left(\frac{x}{x_0}\right)^2\right] \, .
\label{formf}
\ee
Let us remind that the solution obtained for $C(r,t)$,
given in Eq.~(\ref{scalAnsatzSimple}), with $f(x)$ and $L(t)$ given in 
Eqs.~(\ref{formf}) and~(\ref{growthagt4}), respectively,
is only valid for small values of $x$, namely
for $x<x^*$. As we will show at the end of this
section, $x^*$ is a weakly increasing function of time (see Eq.~(\ref{xstar})).

\begin{figure}[htbp]
	\centering
	\includegraphics[width=0.8\textwidth]{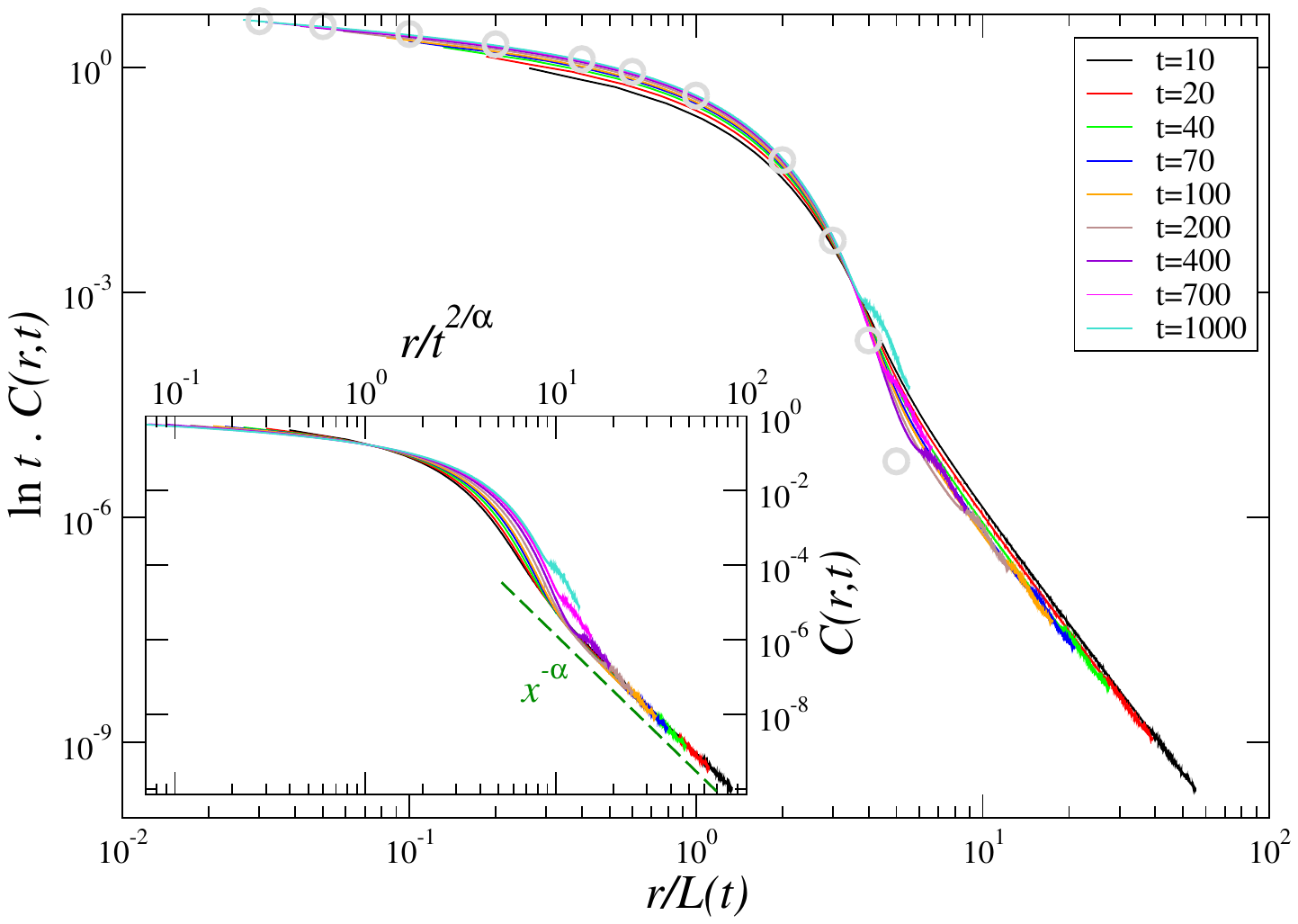}
	\caption{$\ln t\cdot C(r,t)$ is plotted against $r/L(t)$ for $\alpha=5$ at different times (see legend), in a log-log plot. The number of agents is $N=301^2$. The heavy gray circles are the form~(\ref{solutions}) with fitting parameters $c_1=0.55, c_2=0, x_0=1.65$.
		In the inset we plot $C(r,t)$ against $x=r/t^{2/\alpha}$, still with
		double logarithmic scales. The green dashed-line
		is the form $x^{-\alpha}$ of Eq.~(\ref{largexsol1}), with $x=r/\Lambda(t)$.}
	\label{fig_Cagt4}
\end{figure}

The above determinations are compared with
the outcome of the numerical solution of Eq.~(\ref{eqc1}) in Fig.~\ref{fig_Cagt4}, for $\alpha=5$. Because of the form~(\ref{scalAnsatzSimple}), one should find data-collapse of the curves 
at different times (corresponding to different colors in the figure)
by plotting $\ln t \cdot C(r,t)$ against $x=r/L(t)$. This is the kind of
plot presented in the main part of the figure, where $L(t)$ has been extracted from $C(r,t)$ by means of Eq.~(\ref{eqL}). A good data-collapse
can be appreciated, at sufficiently long times, in the small-$x$ sector, for $x<x^*$, where $x^*$
can be located where the master-curve $f(x)$ changes its shape from concave to
straight, in this double logarithmic plot. 
The quality of the data collapse is poorer for short times,
but it constantly improves as time elapses.
The concavity of $f(x)$ in the region $x<x^*$ signals
that the decay is faster than any power, as indeed our analytical
form~(\ref{formf}) implies. 
Actually, in the figure, with gray circles we also plot the
scaling function in Eq.~(\ref{solutions}), where $c_1=0.55$, $c_2=0$, $x_0=1.65$ have been used as fitting parameters. This form interpolates the data for $x<x^*$ very well.
The fact that we have to use $c_1<1$ to superimpose the curve to the data is probably due to the fact that the analytical determination $c_1=1$, $c_2=0$ is only correct
in the very large-time limit when $x^*(t)\gg 1$, as we will comment further below.
The form of $C(r,t)$ for $x>x^*$ will be
investigated later, and we will discuss this part of the figure in due time, and the inset as well. 

The form~(\ref{scalAnsatzSimple}) implies a breakdown of standard dynamical
scaling~\cite{bray94,Puriwad09,CRPHYS_2015__16_3_257_0} (i.e., a scaling as in Eq.~(\ref{scaling})), due to the presence of the logarithmic factor. As in
the NN case, this is due to the building-up of thick interfaces around
correlated regions of size $L(t)$. In order to see this, we 
now compute the density of interfaces $\rho (t)$, which is related to
the correlation function between agents at unitary distance as
\be
\rho(t) \ = \ \ha \, \lf[1-C(r=1,t)\ri] \, .
\ee
In our case
\be
\rho(t) \ = \  \lf[1-\frac{1}{\ln t}\, \text{E}_1\left(\frac{1}{x_0^2L^2(t)}\right)\ri] \, .
\ee
For large $t$, again using Eq.~\eqref{e1small}, one gets
\be
\rho(t) \ \sim \ \frac{\ga- \ln\lf(x_0^2D\ \ri)}{\ln t} \, . 
\ee
Hence the amount of interfaces vanishes only logarithmically, whereas
correlated regions grow algebraically as in Eq.~(\ref{growthagt4}),
similarly to what observed
in the NN model \cite{Frachebourg1996,PhysRevLett.87.045701}.

As we commented already, the solution obtained thus far, based on a 
Maclaurin expansion for small $\ell$, is only valid for sufficiently
small values of $r$, namely for $x<x^*$.
In the following, we determine the form of $C(r,t)$ in the large-$r$
domain, which will be found to be quite different, as can also be appreciated in Fig.~\ref{fig_Cagt4}. This will also allow
us to determine $x^*$. 
For large $r$, the sum on the r.h.s of Eq.\eqref{eqc1} is dominated by the contributions at $k \approx j$, i.e. for $d_k \approx 0$, namely $\ell \approx r$ and $\theta \approx 0$. The first term on the r.h.s. is subdominant and can be discarded, so, going again in the continuum, and using the small argument expansion~(\ref{e1small}) one has
 \begin{equation}
\dot C (r,t) \approx  -2\, \frac{P(r)}{\ln t} \,  \int \!\! \dr \ell \,\dr \theta \, \ell \, \lf[\ga+\ln \lf(\frac{d^2(r,\ell,\theta)}{x_0^2 L^2(t)}\ri)\ri] \, \approx   -2 \,\frac{P(r)}{\ln t} \,  \int \!\! \dr \ell \dr \theta \, \ell \, \ln \lf(\frac{d^2(r,\ell,\theta)}{x_0^2L^2(t)}\ri)\, ,
\label{eqcbx1}
\end{equation}
where, in the last passage, we have dropped sub-dominant terms for large $t$. 
 The approximation~(\ref{e1small}) is correct up to $z\lesssim 1$, namely for $d\lesssim x_0L(t)$.
Since in the above approximation 
$d \approx r \theta \approx \ell \theta$, we can integrate 
up to $\ell \approx x_0 L(t)$. By evaluating the integral and
retaining the dominant term for large $L(t)$ one gets
\be
\dot C(r,t)  \ \propto \ P(r) x_0^2 L^2(t) \, . 
\ee
%
%
Integrating the differential equation yields
\be
C(r,t)  \ \propto \  P(r) \, x_0^2 \, t^2 \, , 
\ee
and then one has a scaling form
\be \label{largexsol1}
C(r,t) \ \propto \ \lf (\frac{r}{\Lambda(t)}\ri)^{-\al} \, ,
\ee
with 
\be
\Lambda (t) \ \propto \  x_0^{4/\al}\, t^{2/\al}.
\ee
Notice that for $x>x^*$, $C$ turns from the fast decay of Eq.~(\ref{formf}) to a slower algebraic decrease, as it can be appreciated in Fig.~\ref{fig_Cagt4}. Moreover, the reference length changes from 
$L(t)\propto t^{1/2}$ to the slower one $\Lambda(t)\propto t^{2/\alpha}$.
This can also be seen in Fig.~\ref{fig_Cagt4}, because a good data collapse
in the region $x>x^*$ is only obtained by plotting $C(r,t)$
against $r/t^{2/\al}$, as done in the inset. By the way, in this figure one can also observe some lattice effects in the form of small ripples at
large $r$. This is because, on the square lattice, periodic boundary conditions have a different effect moving along one of the two easy-axis or
in different directions (the boundary is met sooner upon proceeding along the easy axes with respect to going, say, along a diagonal). Clearly, this fact deteriorates the
data collapse. However this effect is pushed to smaller and
smaller values of the correlation as the thermodynamic limit $N\to \infty$ is taken. Notice also that, with
the rescaling presented in the inset, the collapse at 
$x<x^*$ is definitely worse than the one obtained in the main part of the figure. This effect can be magnified 
upon increasing $\alpha$, since the difference between 
the growth-exponent of $L(t)$ and of $\Lambda (t)$ increases. We do not show this here.

The actual value of $x^*$ can be obtained by matching the two 
forms of $C$ in the large and small $r$ sectors. Using the asymptotic form~\cite{abramowitz1965handbook} of $E_1$
\be \label{e1as}
E_1(z) \ \approx \ \frac{e^{-z}}{z} \, , 
\ee
 we find
\be \label{xstar}
x^*(t) \ \propto \ \sqrt{(\al-4) \ln t} \, .
\ee
As $t \to \infty$, $x^* \to \infty$. This justifies the fact that we could take the limit $x \to \infty$ to fix the boundary conditions in Eq.~(\ref{solutions}) despite the fact that $x < x^*$ in that case.
Notice that $x^*\to 0$ for $\alpha \to 4$, implying again that all the above
ceases to be valid for $\alpha \le 4$. This case will be considered in
Sec.~\ref{secalin24}.

\subsection{Consensus time}

We conclude this section by computing the consensus time $T$, namely the time needed to reach a fully ordered
configuration. When this happens, it must be $C(r,t)\equiv 1$, hence we use the criterion 
\begin{equation}
\sum _{p=0} \ \sum _{k=1}^{n_p}C ([\![d_k(r,\ell_p)]\!],t=T)=\sum _{p=0} \ \sum _{k=1}^{n_p}1= N
\label{criterionConsensus}
\end{equation}
to determine $T$, similarly to what is done in~\cite{PhysRevA.45.1067}. 
Notice that in Eq.~(\ref{criterionConsensus}) the summations start from $p=0$, differently from elsewhere. 
Resorting again to the continuum approximation, transforming the sum into integrals we have
\be
2 \pi \, \int_0^{\cal L} \! \! \dr r \, r \,  C(r,T) \ = \ N \, , 
\ee
where 
\begin{equation}
	{\cal L}\propto \sqrt N
\end{equation}
is proportional to the linear size of the lattice.
In Eq.~(\ref{criterionConsensus}) we can send to $\infty $ the upper integration limit, because $C$ is integrable,
according to the expressions~(\ref{scalAnsatzSimple},\ref{formf}).
One has $\int_0^{\cal L} \! \! \dr r \, r \,  C(r,T)\propto L(T)^2/\ln T$, and hence
\begin{equation}
	T \propto N \ln N,
	\label{consensusagt4}
\end{equation} 
to leading order, as in the NN model~\cite{PhysRevA.45.1067}.

\section{Case \boldmath{$2<\alpha \leq 4$}} \label{secalin24}

\subsection{Stationary states without consensus} \label{statain24}

The NN voter model is impacted by the presence of 
metastable stationary states without consensus in $d\ge 3$~\cite{Clifford1973,Holley1975,Frachebourg1996}, whose
lifetime diverges in the thermodynamic limit $N\to \infty$.
The same feature~\cite{corberi2023kinetics} is found in the $d=1$ model with the same type of long-range interactions we are considering here, for $\alpha \le 2$. We show below that stationary
states with similar characteristics are found also in the
present $2d$ case, but now for $\al \le 4$. Starting with the case
$2<\al \le 4$, it is not difficult to get convinced that
the state with correlation function
\be
C_{stat}(r) \ \propto  \ r^{-(4-\al)} \qquad \mbox{for }r\gg 1 ,
\label{statlarger}
\ee
is stable for $N\to \infty$.
According to Eq.\eqref{eqc1}, letting the l.h.s. equal to zero, this function should obey
\be
C_{stat}(r)={\sum _p}P(\ell_p) \sum _{k=1}^{n_p}C_{stat} ([\![d_k(r,\ell_p)]\!])\, .
\label{autoconsstat}
\ee
Moving to the continuum approximation one has
 \begin{eqnarray}
C_{stat}(r) &\simeq& \int _1 ^r\dr \ell \int \!\! \dr \theta \, \ell \, P(\ell) \, C_{stat}( d(r,\ell, \theta))+\int _r ^{\cal L}\dr \ell \int \!\! \dr \theta \, \ell \, P(\ell) \, C_{stat}( d(r,\ell, \theta)\, .
\label{eqccon}
\end{eqnarray}
Since for $r\gg 1$ Eq.~(\ref{statlarger}) holds, while
$C(0)\equiv 1$, we can use the interpolating form
\be
C_{stat}(r)\simeq (1+\kappa r)^{-(4-\al)}
\label{interpstat}
\ee
in the integral, where $\kappa$ is a constant, thus having
 \begin{eqnarray}
  Z \,C_{stat} (r) &\simeq&  \int _1^r\dr \ell \int \!\! \dr \theta \, \ell^{1-\al}\left [1+\kappa \sqrt{r^2+\ell^2-2 \ell r \cos \theta}\right ]^{-(4-\al)}\nonumber \\
  &+&\int _r^{\cal L}\dr \ell \int \!\! \dr \theta \, \ell^{1-\al}\left[1+\kappa\sqrt{r^2+\ell^2-2 \ell r \cos \theta }\right]^{-(4-\al)}\, ,
\label{eqccon1}
\end{eqnarray}
where we used Eq.~(\ref{eqd}).
The integrals are dominated by the region $\theta \simeq 0$, because at this angle there is a peak of the
integrands at $\ell =r$. Letting $\theta=0$ in the integrands, one arrives at
 \begin{eqnarray}
  Z \, C_{stat} (r) &\simeq&  2\pi\int _1^r\dr \ell \,\ell^{1-\al}[1+\kappa (r-\ell)]^{-(4-\al)}
  +2\pi \int _r^{\cal L}\dr \ell \, \ell^{1-\al}[1+\kappa(\ell-r)]^{-(4-\al)}\, .
\label{eqccon2}
\end{eqnarray}
Working now, for simplicity, for ${\cal L}/r\gg 1$ (however, such limitation can be 
relaxed), the second integral is negligible and, integrating the first term, one has
\be \label{int1}
ZC_{stat}(r)\simeq 2\pi \,\frac{(\kappa  (r-1)+1)^{\alpha } (\kappa +(\alpha -3) \kappa  r-1)-((\alpha -2) \kappa -1) (\kappa  (r-1)+1)^3}{(\alpha -3) (\alpha -2) \kappa ^2 r^2 (\kappa  (r-1)+1)^3} \, \sim  \ 2\pi \, \frac{(\kappa \, r)^{-(4-\alpha)}}{\alpha-2} \, ,
\ee
with the last passage holding for large $r$.



Evaluating $Z$ by also transforming the sum in Eq.~(\ref{eqZ}) into an integral as 
\begin{equation}
	Z\simeq 2\pi \int _1 ^{\cal L}d\ell \, \ell^{1-\al},
	\label{Zint}
\end{equation}
for large ${\cal L}$ one finds $Z \simeq 2\pi/(\al-2)$. Then Eq.~(\ref{int1}) reads
\begin{equation}
	ZC_{stat}(r)\simeq Z(\kappa r)^{-(4-\al)},
\end{equation}
 providing consistency to the initially postulated form~(\ref{statlarger}).

\begin{figure}[htbp]
	\centering
	\includegraphics[width=0.8\textwidth]{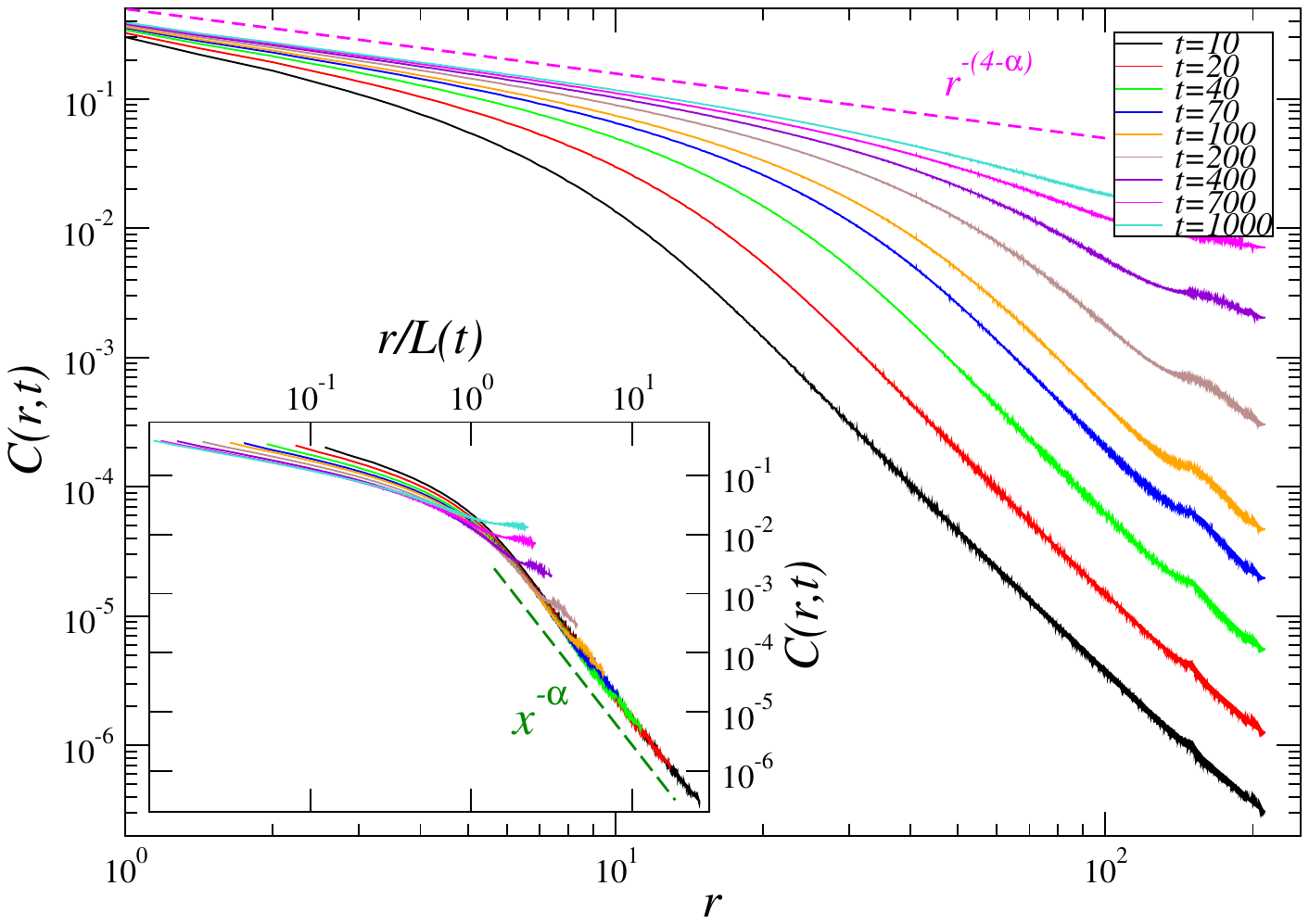}
	\caption{$C(r,t)$ is plotted against $r$, on a double logarithmic scale, at different times (see legend), for $\alpha =7/2$. The number of agents is $N=301^2$. The magenta dashed line is the
	 algebraic behaviour~(\ref{statlarger}).
	 In the inset $C(r,t)$ is plotted against $x=r/L(t)$.
	 The green dashed line is the scaling behaviour~\eqref{scaling},\eqref{scalfunct} in the coarsening stage preceding stationarization.} 
	\label{fig_C_alfa3_5}
\end{figure}

In Fig.~\ref{fig_C_alfa3_5}, the evolution of $C(r,t)$, obtained numerically solving Eq.~(\ref{eqc1}) on the square lattice, is shown for $\alpha =7/2$. In the main part of the figure it can be seen that $C(r,t)$ gradually approaches the stationary form~(\ref{statlarger}) (dashed magenta line), starting from smaller values of 
$r$.

Clearly, the stationary state~(\ref{statlarger}) is not fully ordered, since the criterion~(\ref{criterionConsensus}) is not met. Still,
it is strongly correlated, because the correlation length
$L_{stat}$ diverges in such a state. Indeed, 
using the definition~\eqref{eqL}, one finds 
\be 
L_{stat} \ = \ 
	\frac{\al-2}{\al-1} \, {\cal L}\propto \frac{\al-2}{\al-1}N^{\frac{1}{2}}  \, ,  
\label{Lstatagt2}
\ee
for large $N$. 

\subsection{Consensus time}

The lifetime of stationary states diverges
for large-$N$ but is otherwise finite. Hence,
in a finite system, consensus is always reached
in a limited time $T$.
We saw in Sec.~\ref{secalgt4} that $T\propto N\ln N$ for $\al >4$ (Eq.~(\ref{consensusagt4})). 
We did not find a simple way to compute the consensus time for $\al \le 4$. However,
using the criterion that $T$ must be a non decreasing function of $\al$, because global ordering is promoted by the extent of interactions, and recalling that  
it is $T=N$~\cite{Dben-Abraham_1990} in mean-field (i.e. $\al=0$), we 
conclude that the $N$ dependence of $T$
for $0<\al\le 4$ must be compressed between
$N$ and $N\ln N$. Since for $\al >4$ the $\ln N$ factor is caused by the logarithmic correction to scaling in Eq.~(\ref{scalAnsatzSimple}), and there is no such correction for any value $\al <4$ (as we will see, this is true down to $\al =0^+$) we argue that
$T\propto N$ for any $0\le \al \le 4$.

\begin{figure}[htbp]
	\centering
	\includegraphics[width=0.8\textwidth]{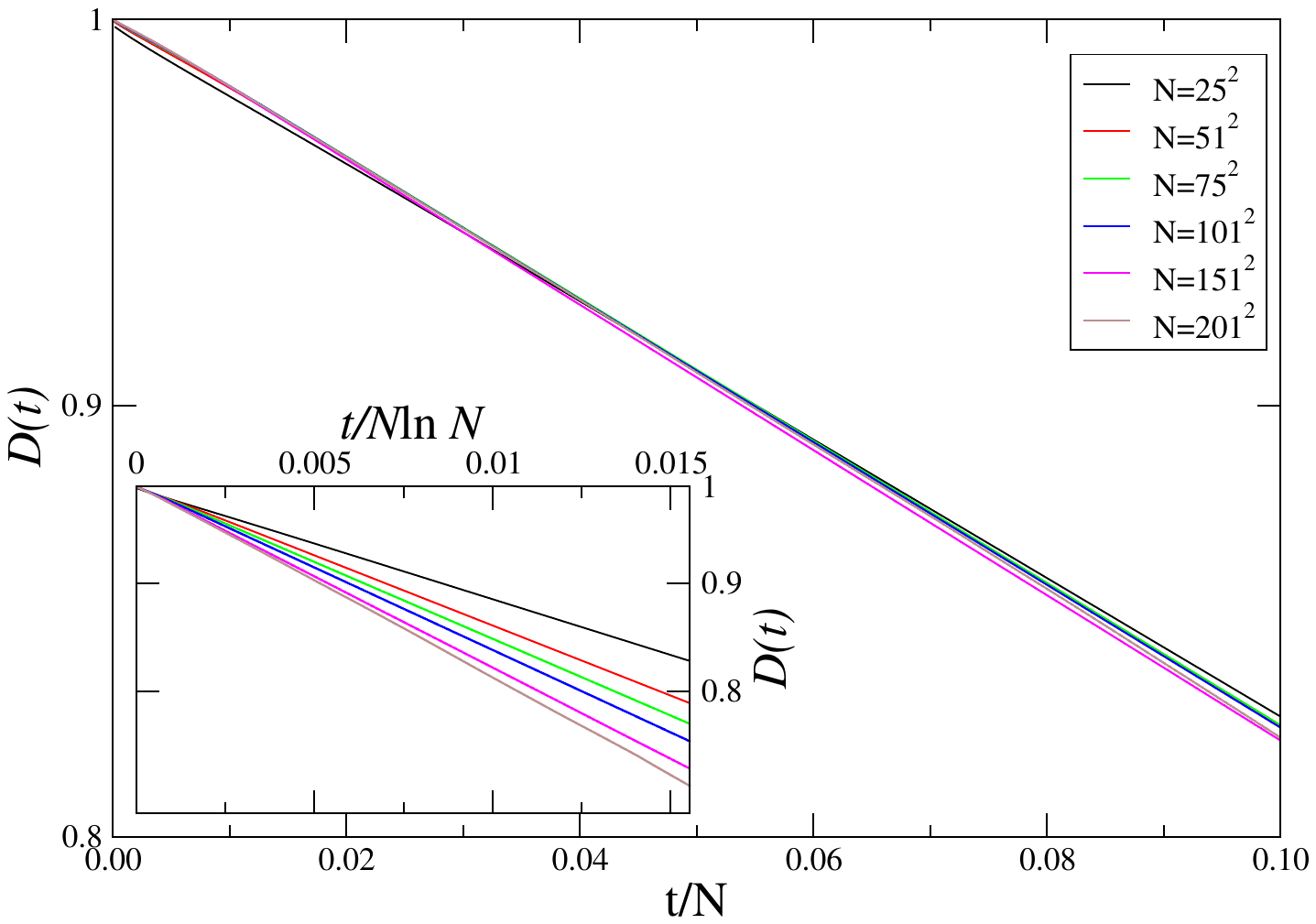}
	\caption{${\cal D}(t)$ is plotted against $t/N$ (main figure) or against $t/(N\ln N)$ (inset). The vertical scale is logarithmic. Different curves correspond to various values of $N$, see legend.}
	\label{fig_consensus_3_5}
\end{figure}

In order to check this hypothesis,
recalling the consensus criterion~(\ref{criterionConsensus}),
we have computed the {\it distance} from consensus
\begin{equation}
	{\cal D}(t)=\sum _pP(\ell_p) \sum _{k=1}^{n_p}1-
	\sum _pP(\ell_p) \sum _{k=1}^{n_p}C ([\![d_k(r,\ell_p)]\!],t).
\end{equation}
In this equation the first term on the r.h.s. is a number that can be exactly computed by enumerating all the possible distances on the lattice, while the second term on the r.h.s. is readily computed once $C$ is known, which is done by means of the
numerical solution of Eq.~(\ref{eqc1}).
Consensus is reached when ${\cal D}=0$.
We find that ${\cal D}(t)$ decays to zero exponentially.
This can be seen in Fig.~\ref{fig_consensus_3_5}, where the case with $\al=7/2$ is considered. Here we plot only a relatively short time interval, because
for longer times numerical errors tend to increase and data become progressively less affordable. From the main figure it is seen that 
curves for different values of $N$ can be collapsed on a single one by plotting against
rescaled time $t/N$. Instead, the rescaling with $t/(N\ln N)$, shown in the inset, fails
in collapsing the curves. Hence we can conclude
that $T\propto N$ is correct.
Summarizing the results for the consensus time
for all values of $\al$ we have

\bea
T(N) \ \propto \ \begin{cases}
	N \ln N & \al >4 \\[2mm]
	N & 0<\al \leq 4 
\end{cases} \, .
\eea

\subsection{Approaching stationarity by coarsening}

We have seen (Eq.~(\ref{Lstatagt2})) that
the stationary states are characterized by a macroscopic correlation length. Since this
quantity starts from a microscopic value, it
must grow, due to a coarsening phenomenon that
will be arrested in the stationary state.
This can be appreciated in Fig.~\ref{fig_length} where a clear increase of 
$L(t)$ is observed for any value of $\alpha$.

In order to compute the properties of such 
pre-asymptotic coarsening 
we have to distinguish between the cases with $3\le \al\le 4$ and $2<\al<3$.

\subsubsection{$3\le \alpha\le 4$} \label{algt3}

We can argue that $C(r,t)$ approaches the stationary form~(\ref{statlarger}) initially for small values of $r$, and then for larger and larger distances. This is very well observed in Fig.~\ref{fig_C_alfa3_5}. 
We will also prove, by checking for consistency at the end of the present calculation, that a large-$r$ scaling form for $C$
\begin{equation}
	C(r,t)=f\left (\frac{r}{L(t)}\right ), \qquad \mbox{for }r\gg L(t)
	\label{scaling}
\end{equation}
is obeyed with 
\begin{equation}
	f(x)\propto x^{-\al},\qquad \mbox{for }x\gg 1,
	\label{scalfunct}
\end{equation}
where $L(t)$ is a function of time to be yet
determined. Looking at the inset of Fig.~\ref{fig_C_alfa3_5}
one can already be convinced, at least visually, that
this is correct. The crossover point $r^*$ between the
small and large-$r$ behaviors can be found by matching the
two forms~(\ref{statClt2}) and~\eqref{scaling},\eqref{scalfunct}, thus obtaining
\be
r^*(t)\propto L(t)^{\frac{\alpha}{2(\alpha-2)}}.
\label{rstar}
\ee
Plugging the scaling assumption~(\ref{scaling}) into Eq.~(\ref{eqc1}) we arrive at
\be \label{diffeqtrue}
\frac{\dot{L}(t)}{L(t)} \,  x \, f^{'}(x) \ = \ 2 \lf[ f(x) -\sum _p P(\ell_p) \sum _{k=1}^{n_p}f\lf(\frac{[\![d_k(r,\ell_p)]\!]}{L(t)}\ri)\ri] \, .
\ee
Once again we have to evaluate the sum appearing in 
Eq.~(\ref{diffeqtrue}). In this case we can restrict
the summation domain to the region $d_k(r,\ell_p)<r^*$, because the decay of $C$ is integrable after $r^*$ and not before.
Transforming the sum into an integral and using the interpolating expression~(\ref{interpstat}) we have
to evaluate
\be
S(r) \ = \  \int \dr \ell \int \!\! \dr \theta \, \ell P(\ell)\left [1+\kappa \sqrt{r^2+\ell^2-2 \ell r \cos \theta}\right ]^{-(4-\al)},
 \label{integ1}
 \ee
on the restricted domain mentioned before.
Working for $r\gg r^*$, the set of allowed values of $\theta$ satisfying these constraints is $\theta \in [-r^*/r,r^*/r]$, i.e., a tiny interval.
Therefore we can safely set $\theta =0$ in the integral and
replace $\int d\theta \to 2r^*/r$, the interval amplitude. Then we arrive at 
\be
S(r) \ = \  \frac{2r^*}{r}\int_{r-r^*}^r \dr \ell \, \ell P(\ell)\left [1+\kappa (r-\ell)\right  ]^{-(4-\al)}+
\frac{2r^*}{r}\int_r^{r+r^*} \dr \ell \, \ell P(\ell)\left [1+\kappa (\ell-r)\right  ]^{-(4-\al)}.
\ee
The algebraic function $\ell P(\ell)$ does not appreciably vary in the integration interval, because we are considering $r\gg r^*$, and hence 
the integrals are approximated by
\be
S(r) \ = \ 2r^*P(r)\int_{r-r^*}^r \dr \ell \,\left [1+\kappa (r-\ell)\right  ]^{-(4-\al)}+
2r^*P(r)\int_r^{r+r^*} \dr \ell \,\left [1+\kappa (\ell-r)\right  ]^{-(4-\al)},
\label{2integ}
\ee
which, for large $r^*$, is
\be
S(r) \ \propto \ \frac{P(r)}{\alpha-3}\,{r^*}^{\alpha-2}\propto \frac{x^{-\alpha}}{\alpha-3}L(t)^{-\alpha/2}.
\label{sr}
\ee
Therefore, Eq.~(\ref{diffeqtrue}) gives
\be 
  \frac{\dot{L}(t)}{L(t)} \, x \, f^{'}(x)  \ \propto \  \frac{x^{-\alpha}}{\alpha-3}\,L(t)^{-\alpha/2}  \, \, ,
  \label{eqden3}
\ee
so one has consistency with the 
$x^{-\alpha}$ behavior and
\be
L(t)\propto t^{2/\alpha}, \qquad 3<\al <4.
\label{growthain23}
\ee

The cases with $\al=3$ or $\al=4$ must be considered separately.

For $\alpha=4$, instead of Eq.~(\ref{statlarger}) one has $C_{stat}\propto (\ln r)^{-1}$.
Therefore, proceeding as before, one finds 
$r^*/(\ln r)^{1/4}\propto L(t)$, in place of Eq.~(\ref{rstar}). Solving for large $L$ gives
$r^*\propto L(t)[\ln L(t)]^{1/4}$.
For $\al=4$ Eq.~(\ref{sr}) becomes
$S(r)\propto P(r){r^*}^2/\ln r^*$. 
Using the value of $r^*$ thus obtained,
one gets
$S(r)\propto x^{-4}L(t)^{-2}[\ln L(t)]^{-1/2}$. Inserting this result into the differential equation~(\ref{diffeqtrue}) and solving for
large times we arrive at
\begin{equation}
	L(t)\propto t^{\frac{1}{2}}(\ln t)^{-\frac{1}{4}}, \qquad \al=4.
	\label{growthal4}
\end{equation}

The behavior for $\al=3$ can also be obtained proceeding along the same lines. In this case the integrals in Eq.~(\ref{2integ}) amount to
\be
S(r) \ = \ \frac{2 L(t)^{3/2}\ln \left(\kappa \, L(t)^{3/2}+1\right)}{\kappa \, r^3} \ = \ 2 \, \kappa^{-1} \, x^{-3} \, L(t)^{-3/2} \, \ln \left(\kappa \, L^{3/2}+1\right) \ .
\ee
Upon substituting this form into Eq.~(\ref{diffeqtrue}),
we again find consistency with the form~(\ref{scalfunct}) and, for large times,
\begin{equation}
	L(t)\propto (t\ln t)^{2/3}, \qquad \al=3.
	\label{growthal3}
\end{equation}

It can be seen in Fig.~\ref{fig_length} that, in the interval $3\le \al \le 4$
the growth-laws predicted in this section are well verified. At very late times 
the curves bend due to the presence of the stationary state and, later on, flatten due to the finite size effect.
Notice that, since $C(r=1,t)=C_{stat}(r)$, in this coarsening stage the density of interfaces $\rho$ stays constant. This holds true for any $\al \leq 4$.

\subsubsection{$2<\alpha <3$}

The $\alpha -3$ at the denominator in Eq.~(\ref{eqden3}) indicates
that the above approach is only valid for $3\le \alpha \le 4$. 
The difference from the previous case is that now, when searching for the solution of Eq.~(\ref{diffeqtrue}), still for large $r$, the restriction $d(r,\ell _p)<r^*$ that we imposed before in Sec.~\ref{algt3}) must be replaced
with a more stringent one, $d_k(r,\ell_p)<\tilde r\ll r^*$, at large times.
The reason is the following: for $\ell \ll r$ the summand ${\cal S}(\ell,r,t)$ in the last 
term of Eq.~(\ref{diffeqtrue}) can be obtained by letting 
$C([[d_k(r,\ell_p)]],t)\simeq C([[d_k(r,0)]],t)=C(r,t)\sim x^{-\alpha}$ ($x=r/L(t)$).
Hence, calling ${\cal S}_<$ the behavior of ${\cal S}$ at small $\ell$, one has ${\cal S}_<(\ell,r,t)\sim x^{-\alpha}P(\ell)$. Instead, for $\ell_p \sim r$ one has 
${\cal S}(\ell_p,r,t)\simeq {\cal S}_>(\ell_p,r,t)\simeq P(r)C(d_k\simeq 0,t)\simeq P(r)[1+\kappa d_k]^{-(4-\alpha)}$, where we again used the interpolating form~(\ref{interpstat}). 
If ${\cal S}_<(\ell_p=r^*,r,t)\ll {\cal S}_>(\ell=r^*,r,t)$ one can proceed as in the previous section~\ref{algt3}, restricting the sum in Eq.~(\ref{eqc1}), namely the integral~(\ref{integ1}), to $d(r,\ell)<r^*$, as we did. This is correct for sufficiently
large values of $\alpha$, which happens to be the case when the solution presented in the previous Sec.~\ref{algt3} is valid, namely for $\alpha >3$. However,
for $\alpha \le 3$ this is no longer true and the sum in Eq.~(\ref{eqc1}) must be
restricted to distances $d_k<\tilde r$, where $\tilde r$ is defined as the distance
where ${\cal S}_<(\ell_p=\tilde r,r,t)\simeq {\cal S}_>(\ell_p=\tilde r,r,t)$. We have not found 
a simple way to explicitly determine $\tilde r$. For this reason, letting 
$\tilde r\sim L^\beta$, where $\beta(\alpha)$ is an unknown exponent, we make the simplest possible ansatz that $\beta $ is a linear function of $\alpha$ 
such that $\tilde r \propto r^*$ for $\alpha \to 3$, namely $\beta =a(\alpha -3/2)$,
where $a$ is a constant yet to be determined.
Evaluating the integrals in Eq.~(\ref{2integ}), with the replacement $r^*\to \tilde r$,
for large $\tilde r$ one has 
\be
S(r)=\frac{P(r)}{3-\alpha}\, \tilde r.
\ee
Plugging this result into Eq.\eqref{eqc1}, one finds
consistency with the $x^{-\alpha}$ behavior for $a=1$, namely $\beta=\alpha -3/2$. Asking for the explicit time-dependence to drop out one finds
\be
L(t)\propto t^{2/3}.
\label{growthalt3}
\ee
We can see in Fig.~\ref{fig_length} that this behavior is very well observed for any $\alpha<3$.

In general the growth exponent $1/z$ -- defined by $L(t)\propto t^{1/z}$ -- is a non-decreasing function
of the range of interactions, because ordering is promoted
by far reaching communications. This is what we find here also, since $1/z$ increases from the NN value
$1/2$, for $\alpha >4$, up to $2/3$ at $\alpha =3$. 
The saturation of this exponent to the NN value upon crossing a {\it critical} value $\alpha _{NN}$ of $\alpha$ is quite generally observed. Besides the case at hand, where $\alpha_{NN}=4$, it is also found in the same model in $d=1$, with $\alpha _{NN}=3$ in that case~\cite{corberi2023kinetics}, and also in ferromagnetic models with algebraic interactions~\cite{PhysRevE.49.R27,PhysRevE.50.1900,PhysRevE.99.011301,Corberi_2017,Corberi2019JSM,PhysRevE.103.012108,Corberi2023PRE}.
On the opposite side of the $\alpha$ variation interval,
it is also generally found that $1/z$ saturates to a maximum
value $(1/z)_{max}$ below a certain value $\alpha_{LR}$ of $\alpha$.
Here we find $\alpha _{LR}=3$, in the corresponding 
$d=1$ voter model it is found~\cite{corberi2023kinetics}
$\alpha _{LR}=2$, while for the $d=1$ Ising model one has 
$\alpha_{LR}=1$~\cite{CLP_review,CORBERI2023113681}. In the $1d$ voter and Ising models the saturation value is $(1/z)_{max}=1$~\cite{corberi2023kinetics,CLP_review}, and this has a simple interpretation. Indeed, the motion of interfaces can be thought of as an advection-diffusion process, where advection is due to the long-range interactions and diffusion to thermal fluctuations or other noise sources. When interactions are extremely long-ranged, for $\alpha <\alpha_{LR}$, the advection fully determines the interfaces motion, leading to a ballistic behavior with
$1/z=(1/z)_{max}=1$. What is interesting in the present
$2d$ voter case, is the appearance of a non trivial value
$(1/z)_{max}=2/3$ eluding an interpretation in terms of a fully advected process. It is clear, however, that such exponent results from the competition between the two contrasting effects induced by the same long-range interactions. Indeed, besides promoting global ordering, interaction with far-apart agents at distances larger than the size of the ordered regions also, has a disordering effect. Let us mention, to conclude this short discussion, that a non trivial value $(1/z)_{max}=3/4$ of the maximum growth exponent is also observed in the $2d$ Ising model with analogous long-range interactions~\cite{PhysRevE.103.012108,PhysRevE.103.052122}, although for different reasons of a geometrical nature. 

\section{Case \boldmath{$0<\al \leq 2$}} \label{secallt2}

We show below that the stationary solution
\bea
C_{stat}(r) \ \propto \ 
      r^{-\al},  \qquad \mbox{for }r \gg 1\, 
   \label{statClt2}
\eea
exists.

%

To verify the consistency of this expression we proceed as in the case  $2<\al \leq 4$. We use the interpolating expression
\begin{equation}
	C_{stat}(r)\simeq (1+\kappa r)^{-\alpha},
	\label{interpalt2}
\end{equation}
in Eq.~(\ref{eqccon}).
Working, as in the case of Sec.~\ref{statain24}, for $1 \ll r \ll {\cal L}$ the relevant contribution to the integrals comes from the region $\ell \approx r$ and $\theta \approx 0$, so we can bring the probability $P$ out of such integrals,
thus arriving at
\be
 Z \, C_{stat}(r) \ \simeq 2\pi\, r^{-\al} \, \int_1^r \dr \ell \, \ell \left [1+\kappa (r-\ell)\right  ]^{-\al}+
2\pi \, r^{-\al} \, \int_r^{\cal L} \dr \ell \, \ell \left [1+\kappa (\ell-r)\right  ]^{-\al} \, \simeq 
2\pi \,\frac{\kappa^{-\al} \, {\cal L}^{2-\al}}{2-\al}\ r^{-\alpha}\, .
\label{consalt2}
\ee
Evaluating $Z$ as in Eq.~(\ref{Zint}) one has $Z =\frac{2\pi}{2-\al}\, {\cal L}^{2-\al}$, and hence the result~(\ref{consalt2}) is consistent with Eq.~(\ref{statClt2}). 

Using the definition~\eqref{eqL}, the correlation length
in the stationary state diverges, for large-$N$, as 
\be 
L_{stat} \ = \ 
\frac{\al-2}{\al-3} \ {\cal L}\propto \frac{\al-2}{\al-3} \ N^{\frac{1}{2}} \, .
\ee
Notice that for $\al=2$ we have a logarithmic correction $L_{stat} = {\cal L}/\ln {\cal L}\propto 
N^{1/2}/\ln N$.

Also in this case, therefore, there is a 
macroscopic correlation length at stationarity, implying that some coarsening
must initially characterize the kinetics as can also be appreciated in Fig.~\ref{fig_length}.  
However the time-domain where $L(t)$ grows
does not increase with $N$ and, hence, this phenomenon cannot
be macroscopically observed. This is because,
as we will discuss in the next section, $L(t)$
carries an $N$-dependence (for fixed $t$) when
$\alpha \le 3$. As can be seen in 
Eq.~(\ref{Lsizedepend}), in fact, $L(t)$
is already of the order of $N^{1/2}$ at fixed $t$,
and hence $L_{stat}$ is approached by $L(t)$ in
a time of the order of one. For this reason, we do not
proceed to the determination of the growth-law
for $\alpha<2$ here. However, it is quite natural to expect that the growth-exponent stays equal to its maximum value $(1/z)_{max}=2/3$ for $\al \le2$ also, as seems to be suggested by Fig.~\ref{fig_length}, even if the power law behavior is too limited in time, because of the reasons explained just above, to arrive at more definite conclusions.

\section{Size dependence of the coarsening domains} \label{secsizedep}
For $\al >4$ we can use the scaling-ansatz \eqref{scalAnsatzSimple} and the definition \eqref{eqL}, employing the continuum approximation, to compute the $N$ dependence of $L(t)$
\be
\frac{\int^{\frac{{\cal L}}{ L}}_{\frac{1}{L}} \!\! \dr x \,  x^{2} \, f(x)}{\int^{\frac{{\cal L}}{2 L}}_{\frac{1}{L}} \!\! \dr x \,  x \, f(x)} \ = \ 1 \, .
\ee
Note that the $\theta$ dependence and the $1/\ln t$ factor are the same for the numerator and the denominator and can be canceled. The integral can be easily performed noting that the form for large $x$ is the one that is relevant to understand the $N$ dependence. Then one finds that 
\be
L(t) \ = \ \frac{(\alpha -2) \left({\cal L}^{\alpha }-{\cal L}^3\right)}{(\alpha -3) \left({\cal L}^{\alpha }-{\cal L}^2\right)} \, . 
\ee
Then $L $ does not depend on $N$ for $N \gg 1$. Following the same procedure, one can check that such behavior still holds for $3<\al \leq 4$. In the limiting case $\al=3$, one has $L \propto \ln N$. For $2<\al<3$ it is $L \propto N^{\frac{3-\al}{2}}$, while for $N=2$, one gets $L \propto \sqrt N/\ln N$. Finally, for $\al <2$, $L \propto \sqrt N$. Summarizing
\be
L(t) \ \propto \left \{ \begin{array}{lcl}
\frac{\alpha -2}{\alpha -3}  \, \mathcal{L}^0 \ \propto \ N^0, &\qquad \mbox{for } & \al >3\,, \\[2mm]
\ln \mathcal{L} \ \propto \ \ln N, &\qquad \mbox{for } & \al = 3\,, \\[2mm]
\frac{\alpha -2}{3 -\al}  \, \mathcal{L}^{3-\al} \ \propto \  N^{(3-\al)/2}, &\qquad \mbox{for } & 2<\al<3\,, \\[2mm]
\mathcal{L}/\ln \mathcal{L} \ \propto \  \ \propto \  \sqrt N/\ln N, &\qquad \mbox{for } & \al =2\,, \\[2mm]
\frac{\alpha -2}{\alpha -3}  \, \mathcal{L} \ \propto \  \sqrt N, &\qquad \mbox{for } & \al <2 \,.\\[2mm]
\end{array} \right .
\label{Lsizedepend}
\ee

\begin{figure}[htbp]
	\centering
	\includegraphics[width=0.8\textwidth]{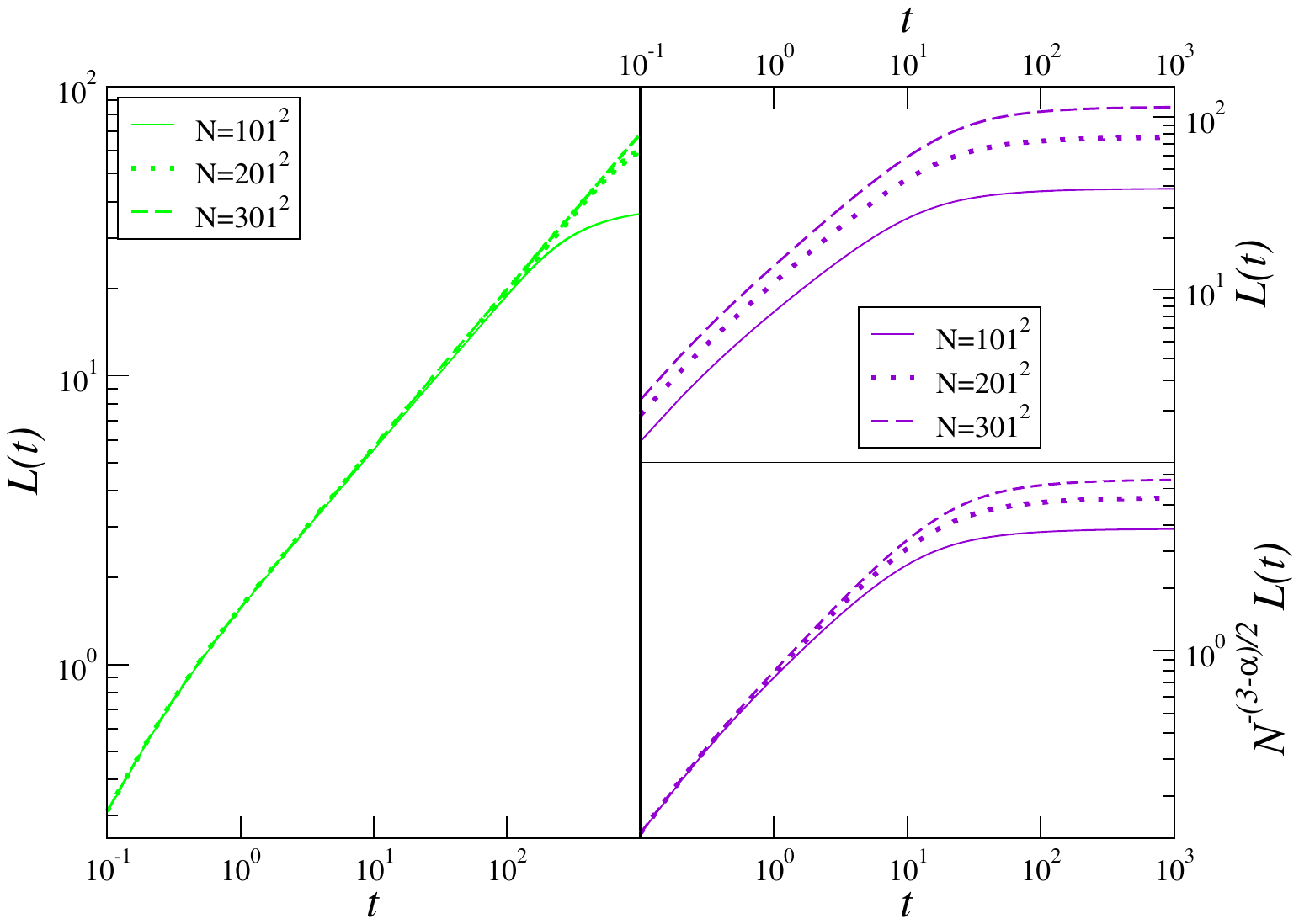}
	\caption{In the left part of the figure $L(t)$ is plotted against $t$ for $\al=4$ and different system sizes, see legend. In the upper-right part the same plot
		is shown, but for $\al =5/2$. In the lower-right part,
		still for $\al =5/2$, we plot $N^{-\frac{3-\al}{2}}\,L(t)$ against $t$.} 
	\label{fig_Ldependence}
\end{figure}

These behaviors are illustrated in Fig.~\ref{fig_Ldependence}. For $\al >3$, which in the figure is represented by the case $\al =4$ on the left half of the plot, the effect of changing $N$ is only manifested at the time when $L(t)$ approaches ${\cal L}$ by a flattening of
the curves. This is the usual finite-size effect observed in a coarsening system with short-range interactions.
Instead, for $\al \le 3$, represented by the case $\al=5/2$
in the upper-right part of the plot, one sees that, on top of the flattening effect discussed before, there is an $N$-dependence at any time. According to Eq.~(\ref{Lsizedepend}), curves for different system sizes 
should collapse by plotting $N^{-\frac{3-\al}{2}} L(t)$,
which is in fact very well observed in the lower-right part
of the figure.

\section{Conclusions} \label{secconcl}

In this paper we have studied analytically the ordering kinetics of the $2d$ voter model with long-range interactions. In this formulation, agents at distance $r$ agree with probability $P(r)\propto r^{-\al}$. The dynamical properties markedly depend on the value of $\al$.
For $\al >4$ we find a behavior analogous to the one observed in the corresponding model with NN interactions, in that coarsening is characterized by the growth-law $L(t)\propto t^{1/2}$ and scaling is violated due to an abundance of interfacial spins whose density vanishes slowly as $\rho (t)\propto 1/\ln t$. A peculiar feature introduced by the long-range interactions is the presence of a 
characteristic scale $x^*(t)$ ($x=r/L(t)$) beyond which a different
behavior, characterized by standard scaling and another growth-law
of correlations is observed. $x^*(t)$ slowly increases in time 
(see Eq.~(\ref{xstar})), eventually relegating this behavior 
to huge distances, and thus eventually banishing it. However this effect produces sizable corrections at finite times. A similar feature was observed in $d=1$~\cite{corberi2023kinetics}. Coarsening ends up in a fully ordered state in a consensus time $T=N\ln N$, as in the NN case.

The situation changes a lot when $\al $ crosses $\al=4$. 
For $\al \le 4$, in fact,
dynamical scaling is fully reinstated and the correlation length increases as $L(t)\propto t^{1/z}$, with an $\al$-dependent exponent 
$1/z=2/\al$ for $3<\al <4$, and with $1/z=2/3$ for $0<\al <3$.
The other important difference with the large-$\al$ case, is that now
the coarsening kinetics does not lead the system to consensus, but to
a metastable state whose lifetime diverges in the thermodynamic limit. Similar stationary states are exhibited by the corresponding $1d$ model for $\al <2$ and by the the NN model in $d\ge 3$ and in the mean-field. Let us also remark that non-equilibrium stationary states are generally observed in various systems with long-range interactions~\cite{Nardini_2012}. In the present model, metastable states are partially ordered with algebraically decaying correlations and an infinite correlation length (in the 
$N\to \infty$ limit).
Finite systems eventually escape this state and reach consensus in a time of the order of $N$, irrespective of $\al$. Another interesting feature, making the long-range case different from the short-range one, is the $N$ dependence of the coarsening length $L(t)$ at
any time for $\al \le 3$ (see Eq.~(\ref{Lsizedepend})), a feature
that is generally observed in coarsening systems with sufficiently
extended interactions~\cite{corberi2023kinetics,CORBERI2023113681}.

Our study can be framed in the more general topic of the evolution far from equilibrium of systems with slowly decaying interactions~\cite{campa2009statistical,book_long_range,DauRufAriWilk} and, more specifically, in the context of 
phase-ordering kinetics with long-ranges~     \cite{PhysRevE.49.R27,PhysRevE.50.1900,PhysRevE.99.011301,Corberi_2017,Corberi2019JSM,PhysRevE.103.012108,Corberi2021SCI,CORBERI2023113681,Corberi2023PRE,PhysRevE.102.020102}. Given the scarcity of analytically tractable models in this field, we believe that our result can provide a 
relevant contribution to the understanding of such problems. Indeed, 
one can argue that the results arrived at analytically in this paper  
could hardly be obtained by means of numerical simulations.
This is because, particularly for $\al \le 4$, most of the relevant
features are encoded into the properties of correlations at relatively large distances, which are shaded
in simulations by the noise. Indeed, the only numerical attempt
to study the long-range voter model that we are aware of is in 
$d=1$~\cite{PhysRevE.83.011110}, while simulation of the Ising model
with long-range interactions suffer from huge noise and finite-size effects for small values of $\al$~\cite{Corberi2021SCI,CORBERI2023113681}. 

The study undertaken in this paper and, for $d=1$, in~\cite{corberi2023kinetics}, open the way to different possible
generalizations. Firstly, Eq.~(\ref{eqc1}) is exact in any dimension
and, therefore, its analysis is also feasible for $d>2$. On the basis of the knowledge gained so far we expect stationary states without
consensus to be present for any value of $\al $ for $d\ge 3$.
However, besides that, the coarsening properties and other features remain to be
studied. Furthermore, in the context of aging systems, the behavior
of two-time quantities, such as the two-time correlation functions, which have not been widely considered so far, are of great interest and could also be investigated analytically. This and other related topics are interesting research lines left for future activity.

\section*{Acknowledgments}

F.C. acknowledges financial support by MUR PRIN 2022 PNRR.

\section*{References}

\bibliography{LibraryStat}

\bibliographystyle{apsrev4-2}

\end{document}